\documentclass[journal]{IEEEtran}
\usepackage[margin=1.0in]{geometry}
\ifCLASSINFOpdf
\else
\fi
\usepackage{hyperref}
\usepackage{color,soul}
\usepackage{graphicx}
\usepackage{cite}
\begin{document}

\title{Towards a Secure and Resilient\\All-Renewable Energy Grid for Smart Cities}

\author{Charalambos~Konstantinou,~\IEEEmembership{Senior~Member,~IEEE}
}

\IEEEaftertitletext{\vspace{-0.2\baselineskip}}

\maketitle

\begin{abstract}
The concept of smart cities is driven by the need to enhance citizens' quality of life. It is estimated that 70\% of the world population will live in urban areas by 2050. The electric grid is the energy backbone of smart city deployments. An  electric  energy  system  immune to  adverse events, both cyber and physical risks, and able to support the integration  of renewable sources will drive a transformational development approach for future smart cities. This article describes how the future electric energy system with 100\% electricity supply from renewable energy sources requires the ``birth of security and resiliency'' incorporated with its ecosystem.
\end{abstract}


%
\IEEEpeerreviewmaketitle

\section{Introduction}

\begin{figure*}[t]
    \centering
    \includegraphics[width = 0.8\textwidth]{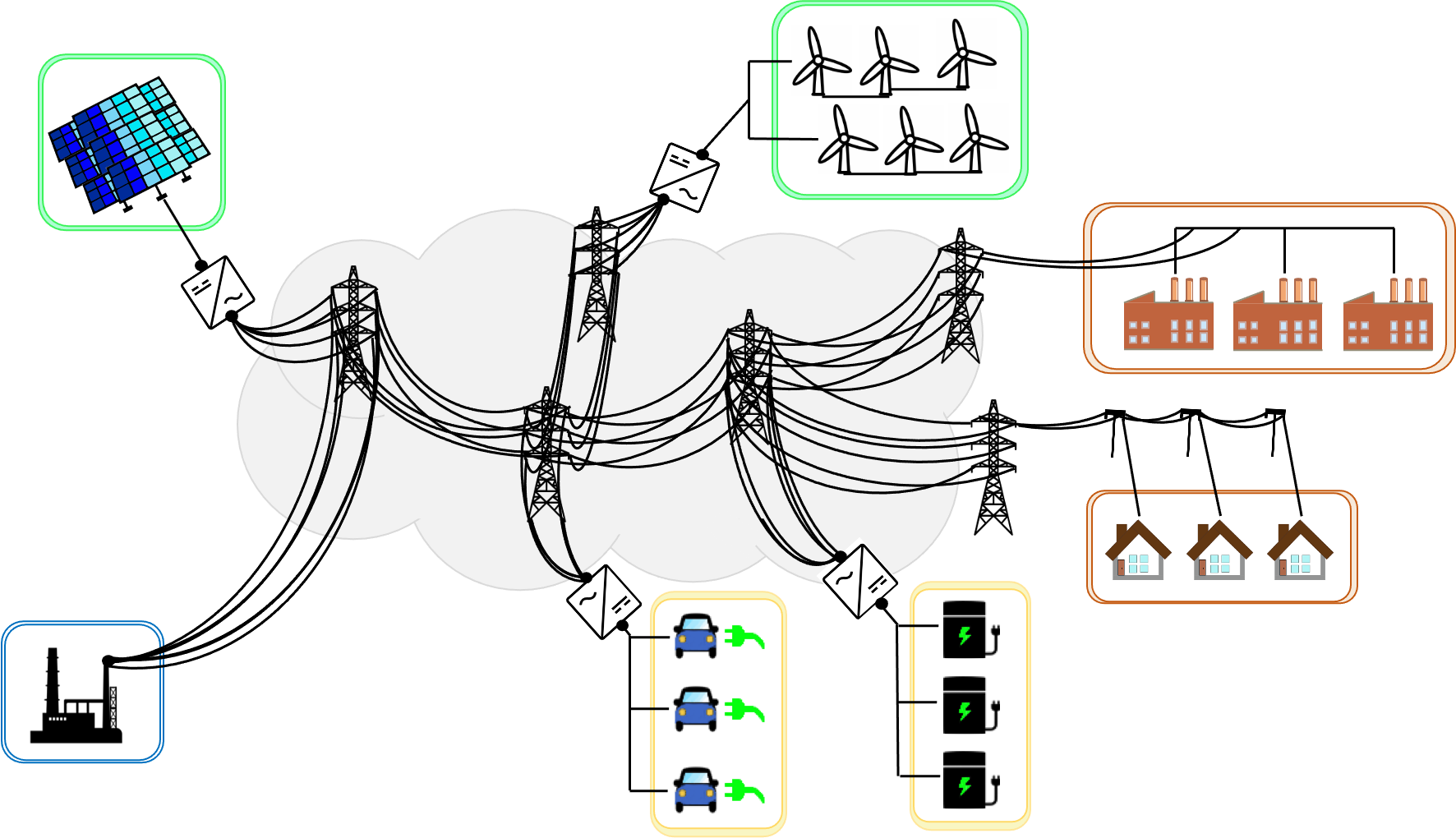}
    \caption{{Simplified diagram of the electric grid integration with renewable energy sources concept.}}
    \label{fig:grid}
\end{figure*}

Electric energy systems 
constitute the backbone of critical infrastructure. National security and economic vitality rely on a safe, secure, and resilient power system. The American electric grid, once considered a marvel of 20th century engineering, has become obsolete in the face of 21st century threats. Our energy grid has numerous shortcomings and can no longer deliver (cyber) secure and (disaster) resilient electric power to businesses and households, 
leading to an urgent and enormous threat to our society and economy. Vertical power systems with rigid transmission and distribution system control hierarchy have failed repeatedly during extreme threats. Recent studies by the Federal Energy Regulatory Commission (FERC) found that knocking out as 9 of the 55,000 power substations could result in U.S. coast-to-coast blackouts lasting 18 months or more \cite{smith2014us}. For example, the Hurricane Michael resulted in 1.7 million power outages along the U.S. Gulf and Atlantic coasts \cite{michael}. During June -- September 2007, heat waves and forest fires occurred in Greece causing extensive damages to the medium-voltage distribution network and knocking out power in many areas of the country \cite{founda2009exceptionally}. Recovery from such disasters also costs tens of billions of dollars including time, manpower, and lost economic productivity, and deepen social inequalities. These failures have taught utilities, regulators, and stakeholders that faults cascade across  national and continental electric grids, and exacerbating a local phenomenon into a socioeconomic catastrophe. Traditional power systems are prone to such cascading power outages that last long periods of time and are complex and time-consuming to recover – in other words, not secure and resilient. Continuing to operate the electric energy system critical infrastructure using the traditional model is a well-recognized security and resiliency threat and the main barrier for the development of future smart cities.

The integration of photovoltaic (PV) solar systems and wind farms together with other renewable energy sources (RES) into the electric grid, as shown in Fig. \ref{fig:grid}, helps towards improving security and reliability of the power system during normal operations and  enhancing resiliency during and after extreme events. In the first quarter of 2018, solar accounted for 55\% of all new generating capacity brought online in the U.S. \cite{eia}, and Florida alone is expected to add over 8.6 GW of solar generation by 2025. 
The inclusion of such distributed resources in the form of solar PV, battery-based storage, and demand resources can increase the resiliency to catastrophic events once research efforts would be able to address open system design questions. Examples include how to strategically locate and operate these resources to sustain smart cities infrastructure by guaranteeing continuity or rapid restoration of power to vital loads following large-scale disturbances by formation of ad-hoc self-contained microgrids in outage situations. In addition, as more and more RES are integrated into power systems, it is projected that offshore oil and gas platforms will be re-used at end-of-life stage for the production of renewable energy (e.g., offshore wind, wave and tidal energy, ocean current energy, ocean-based solar energy, deep-water source cooling, etc.). To thwart the existing problems, {a transformational development approach needs to be established, able to develop and build a secure and resilient electric grid for future smart cities}.  Such development will lead to an electric energy system immune to extreme phenomena while supporting the integration of RES and reducing the dependency on oil drilling into power systems, such as those at the North Sea as well as the Gulf of Mexico and its coastal zone. 

\section{Pillars of a Secure and Resilient All-Renewable Energy Grid}\label{s:attackformulation}

{The objective towards achieving a secure and resilient 100\% renewable energy grid requires the development of multi-layer control and system methods, ranging from system stability controls to secure co-design of hardware/software/physical systems, that contribute towards a holistic cyberphysical energy system (CPES) integrated within smart cities \cite{zografopoulos2021cyberphysical}. The  technological dimensions that will characterize those smart cities are electricity grids and the information and communication networks that can contribute in the development of intelligent transportation systems and support clean energy (e.g., electric vehicles, energy-efficient buildings, etc.).} Technological methods will revolve around decision and cyber-resilient mechanisms of critical CPES and smart cities infrastructures dedicated to advance research and developments on renewable energy modeling and simulation. It is necessary to develop a core innovation ecosystem that enables crosscutting and convergent technology, transformative policy, economic and business models, and develop a future workforce to meet the needs of our cities' future electric grid. Such vision needs to be approached under two main pillars: (1) develop and build a secure and resilient electric grid able to use 100\% renewable energy for electricity, heating/cooling, and transportation and be immune to natural disasters and cyber-attacks; and (2) ensure that the new technologies will be used safely and effectively by operators, and that the new system will be accepted and trusted by consumers and citizens. Our daily lives and economy will be secured, transformed, electrified, and propelled to the next generation. The integration and convergence of these two areas will synergistically catalyze a safer, more reliable, and more efficient community and society. The two directions are described below: 

\subsection{Resilience driven system operation for flexible recovery under extreme events with renewable assets.}

{Smart grids and microgrids operating in conjunction with the bulk power system is an accepted solution and robust in theory \cite{muyeen2017communication, 6172487, 6222538}. There have been multiple smart grid projects which have enhanced one aspect or another to deal with these extreme events -- related challenges -- yet cohesive, comprehensive, general purpose platforms that can be used to drive rapid industry-wide enablement of resiliency are non-existent. Recent surveys in the area clearly indicate the necessity of effective resilience assessment and enhancement methods \cite{arghandeh2016definition, bhusal2020power}.} This direction needs to assimilate advancements in grid modernization technologies achieved through past efforts, modern measurement devices, emerging energy storage and RES  integrations, forecasts, automation, and intelligent algorithms, to create a modernization blueprint for utilities seeking to transform traditional operating architecture to secure and resilient smart energy systems; such that the damage caused by extreme events  is minimized (either cyber-attacks \cite{8634980, ospina2020feasibility}, e.g., Ukraine electric grid attack in 2015 and 2016, weather incidents, e.g., most of the U.S. population live in disaster-prone areas: 75\% of Floridians reside along the coastlines that are routinely threatened by major hurricanes, or even human operation, e.g., 2010 BP oil spill cost is estimated at over \$60 billion and unprecedented damage to the Gulf’s natural environment \cite{bp}).

\begin{figure}[t]
    \centering
    \includegraphics[width = 0.5\textwidth]{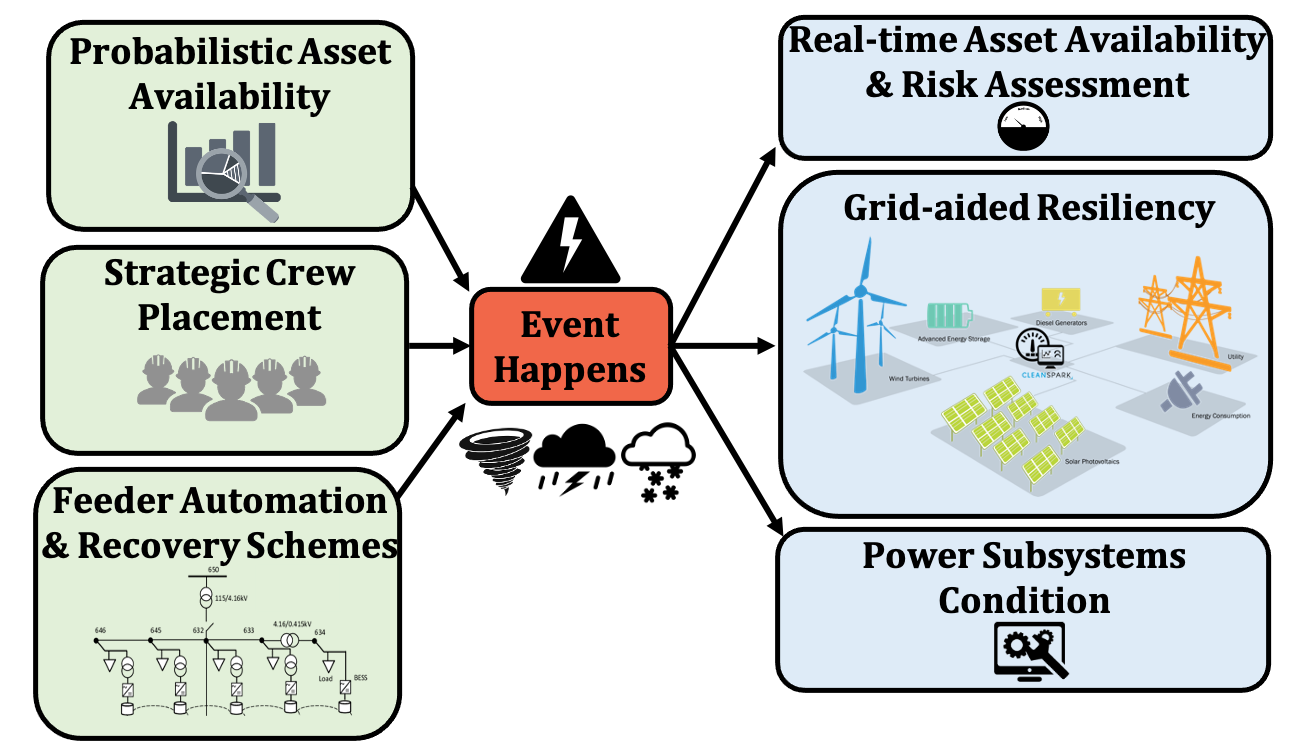}
    \caption{{An example concept of a flexible and agile platform able to adapt to diversifying and expanding threats to power system operation.}}
    \label{fig:platform}
\end{figure}

{This transformation will enable resilient-driven power system operation -- including planning to post-event restoration and recovery -- that would minimize the power outages resulting in less economic losses, public security lapses, and inconveniences. Smart grid frameworks and algorithms need to be encompassed into smart cities and engineered to leverage maturation of microgrid technologies, asset management, and increasing customer-ownership of viable, resiliency-enabling energy resources. 
For example, the smart and automated generation fleet continuity of power supply to critical loads, during and after extreme events, must be driven by intelligent optimization algorithms based on multi-dimensional data resources, probabilistic asset availability forecasting, and resiliency-metrics driven restoration. A graphical illustration of such approach is presented in Fig. \ref{fig:platform}.
The goal from a technology development perspective should be to create flexible and agile platforms that can adapt to diversifying and expanding threats to power system and smart cities operations, and offer just-in-time and post-event intelligence that would aid optimally {and securely} utilities and smart cities units restoration and recovery management processes, i.e., minimizing downtime, operational risks, further exposure to vulnerability, and cost incurred due to the event. Such platforms will need to encompass bulk power systems and distribution feeder-level control, self-contained community microgrid formation with multiple points of common coupling to the surrounding distribution grid, support integration of a fleet of multiple PV systems (able to provide power to critical loads during grid outages), and behind-the-meter resources and customer-owned RES and storage.}

{The development of both planning-based and operation-based platforms supporting RES inclusion requires, not only the incorporation of resilience enhancement strategies, but also needs to encompass  cybersecurity practices and standards able to mitigate growing threats. Existing approaches to mitigate attacks mainly focus on two aspects: (1) resilience control and estimation strategies by excluding the attacked part and adapting optimized control and estimation strategies under attacks,  and (2) attack detection by monitoring unexpected changes in the measured system variables. Attack detection approaches are further classified into two main categories: 
pattern recognition, i.e., learn the recorded historical data corresponding to attacks and then recognize the class of real-time data, and anomaly detection, i.e., compare the real-time behavior of the system with the expected normal behavior.}

{A number of agencies and organizations such as the National Institute of Standards and Technology (NIST), the International Organization for Standardization (ISO), 
and other standards bodies are working together along with the research community to understand new cybersecurity threats, how to create a base set of controls to manage them, understand the risks they pose, and how to mitigate those risks \cite{gritzalis2019critical}. NIST created NIST SP 800-82 \cite{stouffer2014nist}, 
to aid in the implementation of cybersecurity measures in industrial control  systems. 
Also, NIST created two inter-agency reports on internet-of-things (IoT) devices, IoT device cybersecurity capability core baseline (NISTIR 8259A) 
and considerations for managing IoT cybersecurity and privacy risks (NISTIR 8228). 
In addition, utility regulations have cybersecurity requirements and standards. The North Atlantic Electric Reliability Corporation (NERC) published the NERC CIP v5 Cyber Security Standards. The private sector, and specifically natural gas and oil companies \cite{ongscc}, utilize NIST's cybersecurity frameworks and additional international cybersecurity standards including ISA/IEC 62443 series on industrial automation and control systems security, and the Department of Energy (DOE) Cybersecurity Capability Maturity Model (C2M2). Other efforts include NIST's National Cybersecurity Center of Excellence (NCCoE) which works with various critical infrastructure sectors such as electrical utilities and  technology collaborators to explorer the status of cybersecurity concerns that may arise from distributed RES cyber-interconnections \cite{zografopoulos2020derauth}, and develops reference architectures to address these vulnerabilities \cite{nccoe}. Example areas of concern are malware protection and detection, data integrity, device and data authenticity.} 

\begin{figure*}[t]
    \centering
    \includegraphics[width = 0.8\textwidth]{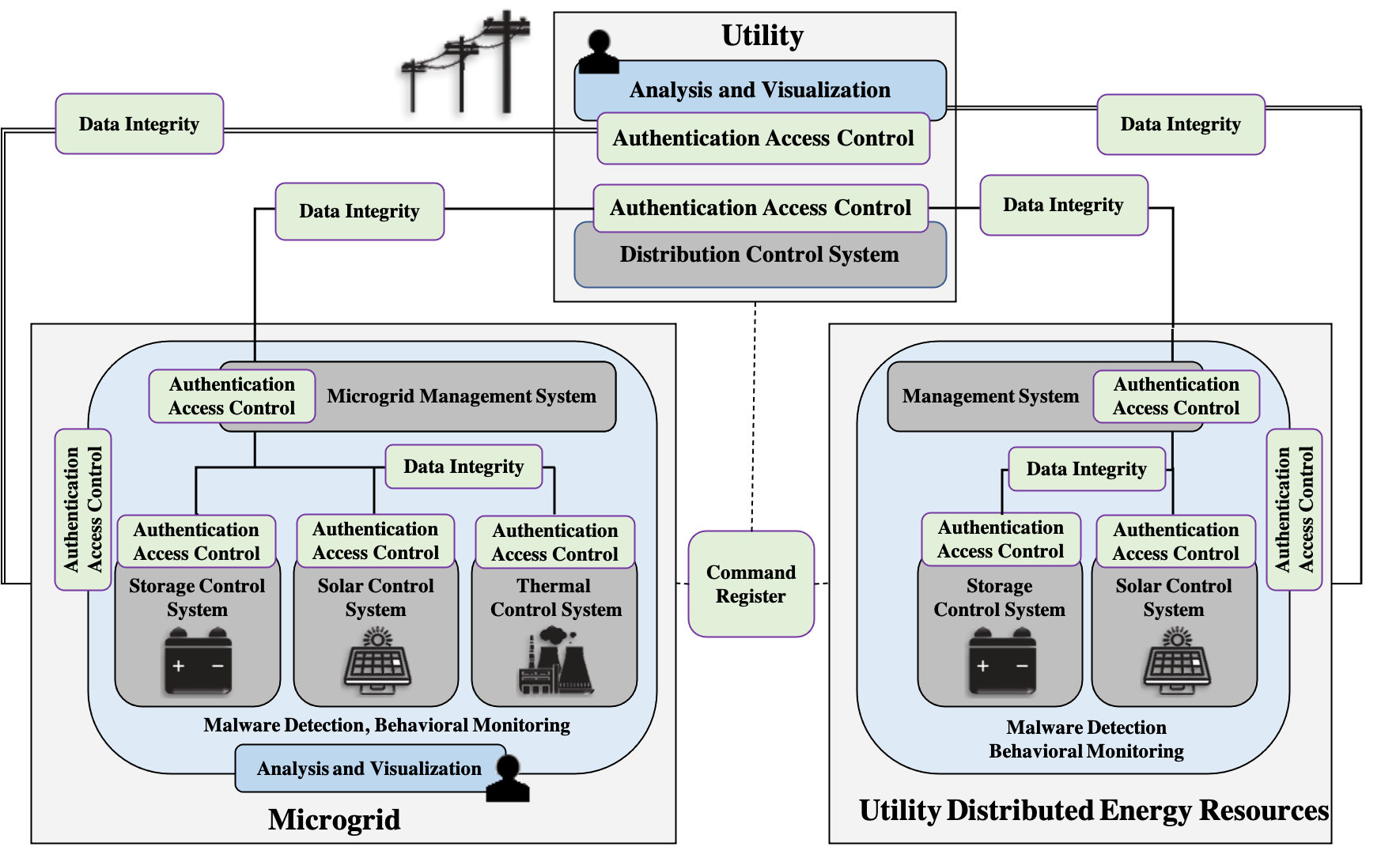}
    \caption{{Cybersecurity capabilities deployed in utility and microgrid infrastructures \cite{nccoe}.}}
    \label{fig:nccoe}
\end{figure*}

{Based on security requirements/outcomes of different attack/concern scenarios and standards/best-practices, cybersecurity capabilities need to include analysis and visualization, authentication and access control, behavioral monitoring, command registers and data historians, data integrity, and malware detection \cite{li2017cybersecurity, kuruvila2020hardware, konstantinou2020cybersecurity}.  
Fig. \ref{fig:nccoe} shows how the desired cybersecurity capabilities may be deployed to protect not only utility systems and distributed RES, but also microgrid and behind-the-meter resources and loads \cite{nccoe}.}

\subsection{Development of an innovation ecosystem for future oil drilling free and all-renewable energy grid.}

Climate change, corporate social responsibility, falling renewable costs, and the drive to diversify have renewed many oil and gas companies’ interest in the renewable energy sector. In addition, the growing energy demand drives RES usage to keep increasing. For example, the U.S. Energy Information Administration (EIA) indicates that, in 2019, wind was responsible for generating approximately 42\% of RES generated power at utility-scale facilities in the U.S., and 7.3\% of the total U.S. electricity generation, making it the most popular RES \cite{eia}.

Although RES integration aids in accommodating an increasing power demand, their volatile intermittent nature causes a mismatch between supply and demand. {Traditionally, alternating current (AC) electric power systems rely on synchronous generators to provide grid stability, and thus achieving an all-renewable energy grid will require to maintain the system stability. The replacement of synchronous machines with PV and wind systems will affect the grid stability since synchronous generators inherently provide the inertial response to power systems. For example, short-term wind power fluctuations occur on a time-scale during which load balancing methods do not yet operate. Since the power system is mainly governed by synchronous inertia, in order to ensure system stability, certain aspects of the electric grid need to be taken into consideration for RES integration (e.g., optimal location, power flow, generation variance, etc.) \cite{liu2020deep}}. 

\begin{figure}[t]
    \centering
    \includegraphics[width = 0.5\textwidth]{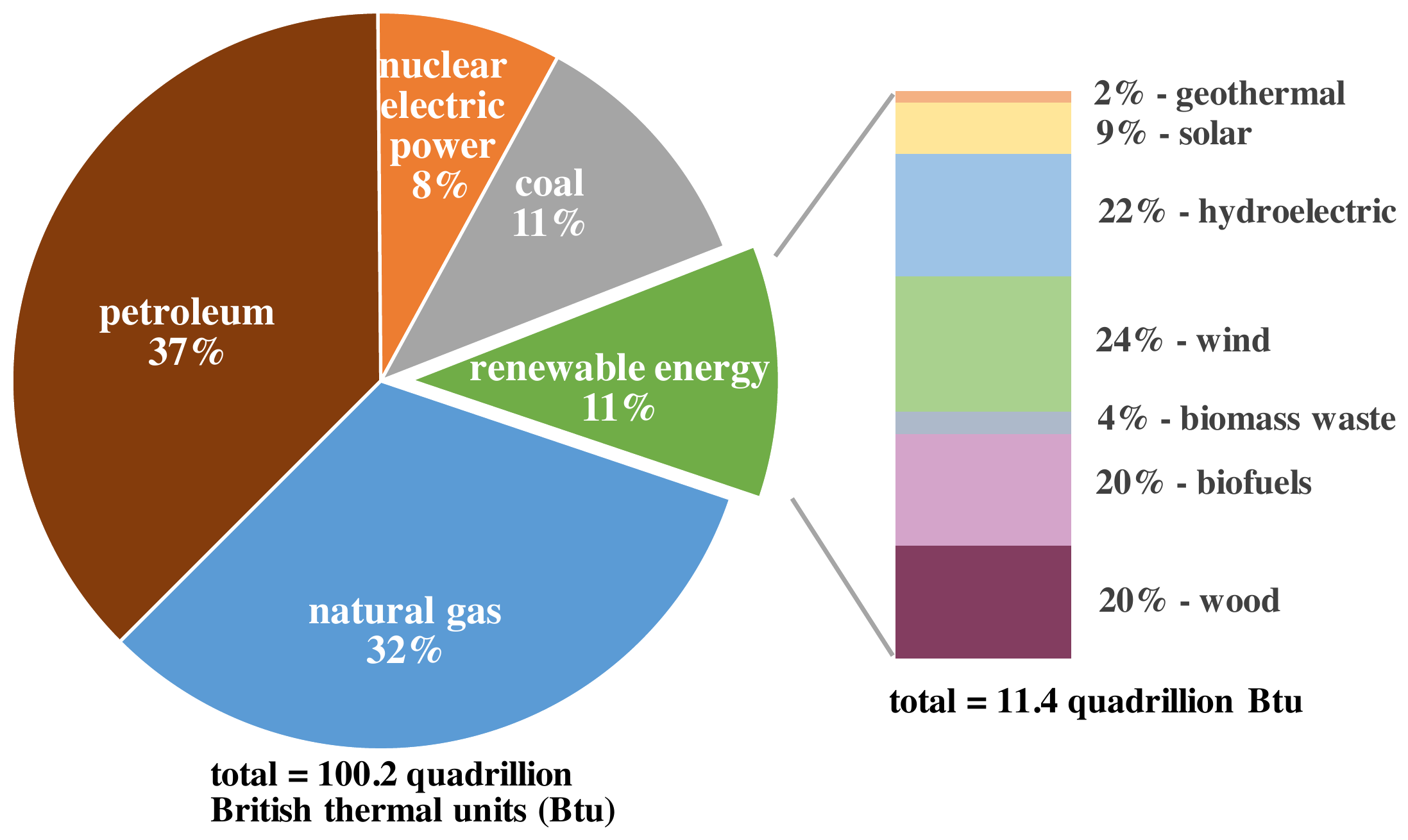}
    \caption{{U.S. primary energy consumption by energy source, 2019. Sum of components may not equal 100\% because of independent rounding. Source: U.S. Energy Information Administration (EIA) 
    \cite{eia}.}}
    \label{fig:piechart}
\end{figure}

{Overall, removing synchronous power generators has the consequent result of less system inertia with impacts on transient and small-signal stability. 
However, if correctly designed, active power controllers for RES within the future inverter-dominated grid can supply a synthetic inertia response to stabilize frequency deviations and maintain grid synchronization. In addition, solar PV systems bolstered by energy storage solutions such as battery energy storage systems (BESS), fuel cells, and flywheels, or derated from their maximum available power capability, can provide synthetic inertia response for under-frequency cases.} While U.S. researchers and firms are driving significant innovations in the RES industry, the U.S. is currently trailing other countries in developing the necessary ecosystem to support the RES integration in the energy sector, and thus, future smart cities \cite{mohanty2016everything, 9194280}. Germany, for example, is a world leader in renewable energy and in the first half of 2018 it produced enough electricity to power every household in the country for a year. For the same year (2018), RES accounted for about 11\% of total U.S. energy consumption {(similar as in 2019 -- 11.4\%, shown in Fig. \ref{fig:piechart})} and about 17\% of electricity generation.

The significance of RES as energy contributor and the opportunities in driving an oil drilling-free ecosystem are not fully understood or recognized by the society as a whole. Consequently, the 
RES industry has suffered from disconnected supply and value chains, and a lack of collaboration. It is essential to structure effective short- and long- term plans for technology development as well as integration and policy growth. Four primary barriers for the innovative ecosystem include: (1) the fragmentation of the 
RES industry for the energy sector, (2) gaps in RES research and manufacturing capabilities, (3)  lacks of business models, public understanding, policies and investment interests, and (4)  workforce availability and talent development challenges. It is critical that governments, 
through public-private partnerships, address these barriers to enable secure and resilient electric grid and capture economic benefit from translating RES research and development into new manufacturing employment and investment. 

\section{Steps Forward}\label{s:methodology}

While solar and other forms of RES are growing rapidly, they are primarily providing off-take energy and, as yet, are largely not integrated into grid operations, nor do they contribute in any significant way to electric and cities systems resiliency. It is critical to effectively utilize RES as a fully integrated grid operating asset and a resource for enhancing security, reliability, and resilience in a substantial and multi-faceted effort. It is important to expand current utility forecasting capabilities not only in terms of assets, but also develop probabilistic models for asset availability, accessibility, and usability during and succeeding a major disaster. The task is complex and is dependent on multiple, interacting and often counter-intuitive correlations between variables. Thus, there is a need to develop new indices for security and resilience, suited for the extreme event challenges of resulting from adverse events, either malicious or not, such as cyber-attacks and  natural hazards including droughts, earthquakes, floods, hurricanes, tornadoes, and fires.

{Future research in the area needs to develop interdisciplinary and innovative technologies for secure and resilient electric grid with 100\% renewable energy vision; put people in the loop in a cyberphysical human-centered systematic approach, formulate and adopt technical, public, and economic policies  to accelerate the use of renewable energy, and develop education programs to train the next generation of power and energy systems engineers. As world urbanization continually grows, smart cities supporting a secure and resilient electric grid will offer an attractive solution that can contribute in economic growth, increased efficiency of energy and transportation systems, and promote sustainable development.}

In terms of anticipated impacts, developments in this area will further lead to: (1) greater clarity and improved understanding of the fundamental gaps in the power system infrastructures and how concerted engineering methods can result in revolutionary improvements in the reliability, efficiency, and resiliency of the fully RES-supported critical city  infrastructure, (2) convene and co-development of the complementary skills needed to address the current and future challenges as well as foster and grow effective relationships with all stakeholders involved,  and (3) preparation of the broader community to adopt the resulting technologies in a timely manner. Societal impact can be achieved by: (1) conducting an inventory of perceived challenges among industry stakeholders to determine prioritized challenges; (2) providing support to industry through advocacy and education; and (3) creating talent pipelines that support industry needs while promoting diversity in the energy sector.

Interdisciplinary environments will facilitate bridging multiple disciplines and viewpoints to enrich the impact of the anticipated research and development as well as  learning experiences. Specifically, it would necessary to draw on the strengths of multiple academic disciplines and collaborations including electrical (e.g., impact of RES generation on the stability of electromechanical oscillation), mechanical (e.g., assessing structural resilience of offshore wind turbines), computer engineering (e.g., defense-in-depth strategies to establish attack-resilient and self-healing security frameworks), social sciences (e.g., uneven recovery after hurricanes has only deepened inequalities based on race and class), and government and industrial stakeholders (e.g., ensure that research outcomes truly represents national needs and urgency). {This broad coalition is necessary to realize a convergent research approach and lead to significant societal impact. 
In terms of education,  interdisciplinary courses merging concepts together from various disciplines will become a priority in order students to acquire more knowledge and skills in the area \cite{9130868}. }

\begin{figure}[t]
    \centering
    \includegraphics[width=0.5\textwidth]{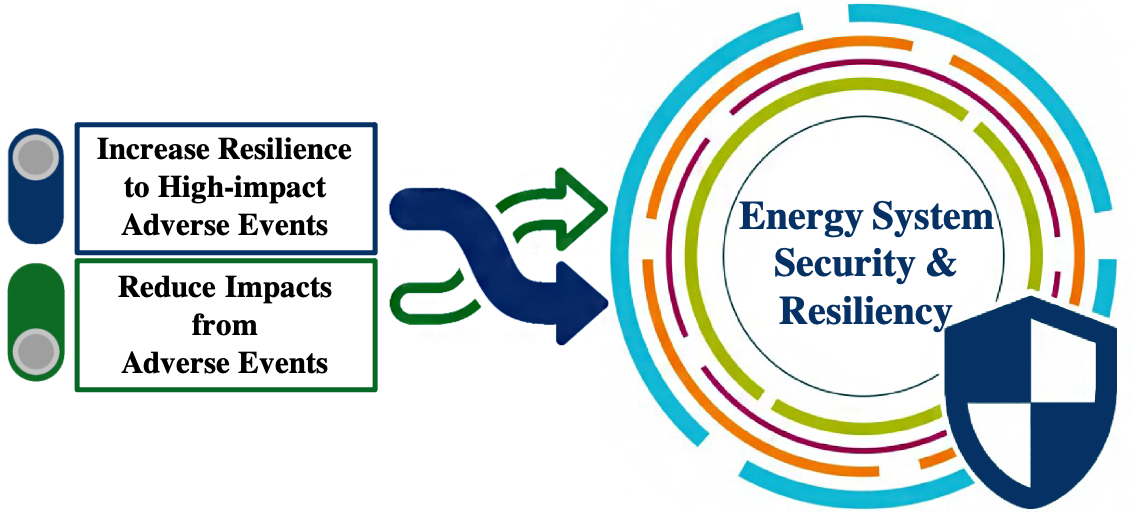}
    \caption{Outcome of future secure and resilient energy systems.}
    \label{fig:concept}
\end{figure}

Looking broadly at the goals of security and resiliency in energy systems utilized in industrial applications and smart city ecosystems, there are two areas where significant progress must still be made, as shown in Fig. \ref{fig:concept}. First, increasing the resilience of critical systems to high-impact low-probability adverse events (e.g., cyber-attacks, weather-induced failures and faults, natural hazards, etc.) is a major priority for CPES and of high significance for safeguarding  national economic and security. Such events can lead, for example, to escalating instability in system dynamics and cause large-area failures. It is of paramount importance to develop resilient and operational-secure strategy methods for electric grid applications in smart city environments considering the renewable future of the oil and gas industry. These methods will take into consideration of offshore installations within the oil and gas industry and how to reuse these structures for RES, leading to a ``green decommissioning'' and to a ``green economy''. Towards this energy transition framework, it is necessary to encounter erratic events as well as the presence of modeling weather uncertainty. As a result, future methodologies would require to address disruptive events by exploiting both localized and supervisory RES schemes, and prevent them from escalating into system failures.

Second, a resilient electric grid should eliminate, or at least significantly reduce, impacts induced from extreme weather- and cyber- based events. Thus, it is important to expand current utility forecasting capabilities not only in terms of assets, but also develop probabilistic models for asset availability, accessibility, and usability during and succeeding a major disaster. The task is complex and is dependent on multiple, interacting and often counter-intuitive correlations between variables. Thus, there is a need to develop new indices for security and resilience, suited for the weather challenges and technological advancements of each smart city. Therefore, in the future, it would necessary  to develop smarter and cost-effective strategies within comprehensive frameworks and algorithms for enabling resiliency of RES-based power systems -- encompassing time-span from planning to post-event restoration and recovery that would minimize the power outages \cite{stright2020defensive}. {A framework based on security and resiliency metrics can be engineered to leverage maturation of microgrid technologies, asset management, conformation to cybersecurity standards and support of state-of-the-art attack detection/prevention methods, and increasing customer-ownership of viable, resiliency-enabling energy resources}.  For example, smart and automated generation fleet continuity of power supply to critical loads, during and after extreme events, can be driven by intelligent optimization algorithms based on multi-dimensional data resources and metrics \cite{8743447}. Adding layers of probabilistic asset availability forecasting can lead to a better demonstration of metrics-driven monitoring and restoration. Security and resiliency metrics will be foundational components of future technology platforms  which would glue together complex and interacting decision variables. This puts security and resiliency of the bulk power system and distribution systems of smart cities at the heart of the modernization challenges.




\bibliographystyle{IEEEtran}
\bibliography{bibliography}

\end{document}